TITLE

# Electron transport in Si/SiGe modulation-doped heterostructures using Monte Carlo simulation

AUTHORS

Florian Monsef [a], Philippe Dollfus [a], Sylvie Galdin-Retailleau [a]

Hans-Joest Herzog [b] and Thomas Hackbarth [b]

[a] Institut d'Electronique Fondamentale, UMR 8622, CNRS, University of Paris-Sud XI, 91405 Orsay, France
[b] DaimlerChrysler AG, Research and Technology, Wilhelm-Runge Strasse, 11, 89081 Ulm, Germany

e-mail: florian.monsef@ief.u-psud.fr
fax: +33 1 69 15 40 20

ABSTRACT

The electron transport in the two-dimensional gas formed in tensile-strained $Si_{1-x}Ge_x/Si/Si_{1-x}Ge_x$ heterostructures is investigated using Monte Carlo simulation. At first the electron mobility is studied in ungated modulation doped structures. The calculation matches very well the experimental results over a wide range of electron density. The mobility typically varies between 1100 $cm^2$/Vs in highly-doped structures and 2800 $cm^2$/Vs at low electron density. The mobility is shown to be significantly influenced by the thickness of the spacer layer separating the strained Si channel from the pulse-doped supply layers. Then the electron transport is investigated in a gated modulation-doped structure in which the contribution of parasitic paths is negligible. The mobility is shown to be higher than in comparable ungated structures and dependent on the gate voltage, as a result of the electron density dependence of remote impurity screening.

I. INTRODUCTION

The use of strained-Si quantum well pseudomorphically grown on a SiGe virtual substrate is becoming a key-factor of channel-engineering to improve the performance of Si-based field-effect transistors (FETs) for both microwave and logic applications. This type of biaxial strain introduces a splitting of degenerate bands [1-2] which results, for both electrons and holes, in smaller in-plane effective mass and reduced intervalley scattering yielding improved transport properties. N-type Si/SiGe modulation-doped FETs (MODFETs) have demonstrated low noise figure and high cut-off and maximum oscillation frequencies [3-4]. The strain-induced enhancement of carrier transport properties is also shown to improve significantly the current in P-channel and N-channel metal-oxide-semiconductor FETs (MOSFETs) designed for CMOS application [5-6].

The starting point of these device concepts is the growth of high-quality Si/SiGe heterostructures taking advantage of the strain-induced mobility enhancement [7-8]. The understanding of their transport properties and of the underlying physics is then a key issue. The aim of this article is to investigate the electron transport at room-temperature in the two-dimensional electron gas (2DEG) formed in $Si_{1-x}Ge_x/Si/Si_{1-x}Ge_x$ heterostructures. The particle Monte Carlo method is used for this purpose. The model includes all relevant scattering mechanisms, i.e. phonon, impurity and alloy scatterings. We first study the electron mobility in ungated modulation-doped structures in which the conducting electrons are provided by either single-side or double-side delta-doped supply layers. The calculated results are carefully compared with measurements for a wide range of electron sheet densities. The mobility in such structures is proved to be very sensitive to the doping parameters influencing the impurity scattering rate, i.e. the doping dose, the position and the shape of the doping profile. These parameters influence the strength of the electron-impurity interaction and the role of possible conducting parasitic paths. To make relevant the comparison with measurements it is thus necessary to overcome the uncertainty on the actual doping profile. An empirical methodology is used to adjust the doping profile in simulated structures which makes them as close as possible to the fabricated samples. Then the crucial influence of the thickness of the spacer layer separating the strained-Si channel from the supply layers is analyzed. We finally study the effect of gate bias on the electron mobility in a gated stack designed for MODFET processing. In this case the pulse-doped layer located between the gate and the strained-Si channel is fully depleted, which suppresses a parasitic path and modifies the overall transport properties. The mobility is found to increase as the gate voltage increases, which is essentially due to the more effective screening of remote doping impurities.

## II. THEORITICAL MODELS

We detail in this section the main features of the Monte Carlo simulator used to calculate the electron mobility in the 2DEG formed in $Si_{1-x}Ge_x/Si/Si_{1-x}Ge_x$ modulation-doped heterostructures.

### A. CONDUCTION BAND STRUCTURE

In this work, the conduction band structure used to model the electron transport in Si and SiGe consists of six ellipsoidal non parabolic $\Delta$ valleys located along the [100] directions at 85% of the *Brillouin* zone edge. The longitudinal effective mass, the transverse effective mass and the non parabolicity coefficient are assumed to be $m_l = 0.9163\,m_0$, $m_t = 0.1905\,m_0$ and $\alpha = 0.5$, respectively, $m_0$ being the free electron mass. The effect of strain is included in the splitting energy $\Delta E_s$ between the two-fold degenerate $\Delta$ valleys (hereafter noted normal valleys) having the longitudinal axis normal to the plane of growth and the four-fold degenerate $\Delta$ valleys (hereafter noted parallel valleys) having the longitudinal axis in the plane. Collecting the results of Refs. [1,9] for the case of tensile-strained Si pseudomorphically grown on $Si_{1-x}Ge_x$ virtual substrate, the splitting energy and the conduction band offset are given by $\Delta E_s = 0.68\,x$ eV and $\Delta E_c = 0.55\,x + 0.1\,x^2$ eV, respectively. The effective masses [10] and scattering parameters are assumed to be unmodified by the strain.

The subband energies in normal and parallel valleys are calculated by solving self-consistently the 1D Poisson/Schrödinger equations within the effective mass approximation. The wave function parallel to the interfaces (*xy* plane) is assumed to be a plane wave and the normalized envelope function in the confinement *z*-direction is denoted $\xi_m(z)$ for the $m^{th}$ subband.

The scattering mechanisms included in the Monte Carlo algorithm are acoustic intra-valley phonon scattering, three *f* and three *g* intervalley phonon scatterings, alloy scattering and impurity scattering. The scattering models are described in the following sub-sections.

### B. ELECTRON-PHONON SCATTERING

The electron-phonon interactions include intra- and inter-valley transitions. All phonon energies and deformation potentials are assumed to be the same as in bulk material and may be found in Ref [11]. The acoustic intra-valley phonon scattering is treated as an elastic process with a scattering rate given by

$$\Gamma_{ac}(E) = \frac{k_B T D_{ac}^2}{2\hbar^3 \rho v^2} \sqrt{m_x m_y} \left(1 + 2\alpha E'\right) \int \xi_n^2(z) \xi_m^2(z) \, dz \quad (1)$$

where $D_{ac}$ is the deformation potential, $\rho$ is the crystal density, $v$ is the sound velocity, $k_B$ is the Boltzmann constant, $T$ is the temperature, $m_x$ and $m_y$ are the effective masses in the $x$ and $y$ directions, respectively, and $\xi_m$ and $\xi_n$ are the envelope functions of the initial and final subbands with energies $E_m$ and $E_n$. The final energy is $E' = E + E_m - E_n$.

Inter-valley transitions are treated by considering three $f$ and three $g$ phonons via either zeroth-order or first-order process in agreement with selection rules [11,12]. For zeroth-order intervalley processes (i.e. for high energy phonons $f_2$, $f_3$ and $g_3$), we use the common model of scattering rate given by [13],

$$\Gamma_{iv_0}(E) = \frac{Z_{iv} D_0^2}{2\rho \hbar^2 \omega_{iv}} \left[N_q + \frac{1}{2} + \frac{\sigma}{2}\right] \sqrt{m'_x m'_y} \left(1 + 2\alpha E'\right) \int \xi_m^2(z) \xi_n^2(z) \, dz \quad (2)$$

where $N_q$ is the average number of phonons at a temperature $T$, $D_0$ is the deformation potential for zeroth-order intervalley process, $Z_{iv}$ is the number of possible final valleys and $\hbar\omega_{iv}$ is the phonon energy. The number $\sigma$ equals $-1$ in the case of a phonon absorption and $+1$ in the case of an emission. Note that prime superscripts refer to final valley. The final energy is $E' = E + E_m - E_n - \sigma \hbar\omega_{iv}$.

For first order inter-valley phonon scattering (i.e. for low energy phonons $f_1$, $g_1$ and $g_2$), we use the scattering rate derived in Ref [14], i.e. by including the non parabolicity of the conduction band

$$\Gamma_{iv_1}(E) = \frac{D_1^2}{\rho \hbar^2 \omega} \left(1 + 2\alpha E'\right) \sqrt{m'_x m'_y} \left[N_q + \frac{1}{2} + \frac{\sigma}{2}\right]$$
$$\times \left(\frac{e}{\hbar^2} \left[\gamma(E)\sqrt{m_x m_y} + \gamma(E')\sqrt{m'_x m'_y}\right] F_{mn} - \frac{1}{2} G_{mn}\right) \quad (3)$$

with,

$$\begin{cases} F_{mn} = \int p_{mn}^2(z)\,dz \\ G_{mn} = \int p_{mn}(z)\,p''_{mn}(z)\,dz \\ p_{mn}(z) = \xi_m(z)\xi_n(z) \quad \text{and} \quad p''_{mn}(z) = \dfrac{d^2 p_{mn}(z)}{dz^2} \\ \gamma(E) = (1 + \alpha E)E \end{cases} \qquad (4)$$

C. ALLOY SCATTERING

In *n*-type Si/SiGe modulation-doped structures the main conduction channel should occur in the strained Si layer. However, in the case of high doping dose in the SiGe supply layers, a parasitic path may be formed in the SiGe alloy. Thus it is necessary to include the alloy scattering mechanism in the simulation.

The classical model of alloy scattering in a 3DEG is based on a "square well" perturbation potential of height $E_{all}$ in a sphere of volume $V$ centered on each alloy site. The radius of this sphere is arbitrarily chosen as the nearest-neighbor distance $r_0 = \sqrt{3}\,a_0/4$ where $a_0$ is the lattice parameter [15]. The alloy potential $E_{all}$ is considered as a fitting parameter equal to 0.8 eV for electrons in SiGe [16].

The spherical symmetry of this standard model is not convenient to derive the transition matrix element between initial and final states in a 2DEG. We propose to slightly modify this model by replacing the spherical region by a cylinder having the same volume $V$, which makes much easier the derivation of the overlap factor between the initial and final envelope functions. To make the cylinder shape close to that of the sphere, we arbitrarily assume the cylinder height $z_1$ to be equal to the diameter $d_1 = 2\,r_1$, that is $z_1 = 2\,r_1 = (2/3)^{1/3}\,r_0$.

The thickness of the 2D structure is discretized into the convenient number of crystalline planes $a_0$ apart. Assuming a random distribution of alloy sites, the derivation of the transition matrix applied to the $p^{th}$ plane at the position $z_p$ gives the following momentum relaxation time $\tau_{all}^{(p)}$

$$\frac{1}{\tau_{all}^{(p)}} = \frac{3^{2/3}\,\pi^2}{2^{11/3}}\,a_0^2\,\frac{E_{all}^2}{\hbar^3}\,x(1-x)\,\sqrt{m_x m_y}\,(1 + 2\alpha E)\,I_p^2 \qquad (5)$$

where $I_p$ is the partial overlap integral defined as,

$$I_p = \int_{z_p-z_1/2}^{z_p+z_1/2} \xi_m(z)\xi_n(z)\,dz \tag{6}$$

By adding up the contribution of all planes the total relaxation time is finally given by

$$\frac{1}{\tau_{all}} = \frac{3^{2/3}\pi^2}{2^{11/3}} a_0^2 \frac{E_{all}^2}{\hbar^3} x(1-x)\sqrt{m_x m_y}\,(1+2\alpha E) \sum_p I_p^2 \tag{7}$$

Each overlap integral $I_p$ is numerically evaluated.

D. IMPURITY SCATTERING

From a modeling point of view, the electron-impurity scattering in a 2DEG gave rise to many works [13,17-19] but there is still a major difficulty that lies in the proper description of screening effects at room-temperature. In the approximation of the linear response of the electron gas, the screening effect is usually introduced in the Coulomb potential through the dielectric function [20]. At 300K, many subbands are occupied and should take part into the screening effects. Siggia and Kwok [21] developed a screening model that takes into account the contribution of several subbands. This approach has been followed by some authors for transport calculation in the approximation of small scattering angles [19,22,23]. A description of multi subband effects has been developed even for device simulation at the price of a complicated and computationally demanding technique [13]. For a forthcoming use at a device scale including quantization effects, we need a simple model of screening function. Therefore we choose a single-subband screening model that will be validated in the next section by comparing computed 2D mobilities with experimental data on a wide range of doping conditions. In many cases the lowest subband of the strained channel is the most populated and gives the correct space extension of the concentration profile. It should be chosen to evaluate the single-subband screening function. This point will be discussed later on.

It has been shown that inter subband matrix elements are negligible in comparison with intra subband counterpart [19]. Therefore, we neglect inter-subband transitions, as most authors do [19,22].

Within these approximations the impurity momentum relaxation time $\tau_{imp}$ for electrons in the $m^{th}$ subband may be written [24]

$$\frac{1}{\tau_{imp}} = \frac{e^4 \sqrt{m_x m_y}}{4\pi \hbar^3 \varepsilon_0^2 \varepsilon_r^2} (1+2\alpha E') \int N_I(z_0) \int_0^\pi \left| \int \frac{\xi_m^2(z) e^{-Q|z-z_0|}}{Q(\theta)+Q_{scr}(\theta)} dz \right|^2 (1-\cos\theta) \, d\theta \, dz_0 \qquad (8)$$

where $N_I(z_0)$ is the ionized impurity concentration at a $z_0$-position, $Q = |\vec{k}-\vec{k}'|$ is the scattering wave vector, $\theta$ is the scattering angle, $\varepsilon_0 \varepsilon_r$ is the dielectric permittivity and $Q_{scr}$ is the screening function. It has been shown that the approach consisting in introducing the relaxation time in the Monte Carlo algorithm is equivalent to the use of the actual scattering rate in terms of computed mobility [25]. Considering a single-subband in the screening effect and including the temperature dependence given by Fetter [26,13] the screening function $Q_{scr}$ is given as a function of $Q$ by

$$Q_{scr}(Q) = \frac{e^2 n_1}{2\varepsilon_0 \varepsilon_r k_B T} g_1(Q\lambda) \iint dz' dz \, \xi_1^2(z) \xi_1^2(z') e^{-Q|z-z'|} \qquad (9)$$

where $n_1$ is the electron density on the screening subband and $g_1(x)$ is defined as

$$g_1(x) = \frac{2\sqrt{\pi}}{x} \Phi\left(\frac{x}{4\sqrt{\pi}}\right) \qquad (10)$$

and $\Phi$ is the dispersion plasma function, defined as,

$$\Phi(y) = 2 e^{-y^2} \int_0^y e^{t^2} dt \qquad (11)$$

For each simulated electron, the subband to be considered in the screening effects must be chosen carefully according to the electron subband and to the relative population of the main and parasitic paths. In Si/SiGe modulation-doped structures a subband may be bound to either the strained Si layer (main channel) or a SiGe supply layer (parasitic path).

For an electron bound to the supply layer, the subband to be considered in the screening function is clearly the lowest subband of the corresponding parasitic path. For an electron bound to the main channel the subband to be chosen is the lowest of the main channel in most cases. In the unfortunate case where the electron concentration in a supply layer would approach the impurity concentration, which is not suitable to get a high mobility, the lowest subband of the corresponding parasitic path is certainly the most effective in the screening mechanism and must be considered for the evaluation of the screening function defined by Eq. (9).

## III. MODULATION DOPED STRUCTURES : MEASUREMENTS AND CALCULATIONS

In this section, we carefully compare transport calculation with mobility measurement in Si/SiGe modulation doped structures. We consider a wide range of layer stacks and doping profiles. We then analyze the influence of the thickness of the spacer layer that separates the strained Si channel from $\delta$-doped SiGe layers. We finally study the electron transport in a gated structure.

### A. STUDIED STRUCTURES

The N-type modulation-doped structures realized for mobility measurements consist in a strained-Si channel embedded in unstrained $Si_{1-x}Ge_x$ layers. Single-side doping (SSD) or double-side doping (DSD) layers are introduced in the epi-layer stack (Sb-doping) to supply the 2DEG with conducting electrons [8]. The cross-section of SSD and DSD structures is shown in Fig. 1. The thickness of the Si channel is always $W_{ch} = 9$ nm, while the spacer thicknesses ($W_{sp_1}$ and $W_{sp_2}$), the nominal doping dose ($W_{\delta_1}, W_{\delta_2}, N_{\delta_1}, N_{\delta_2}$) may vary from a structure to another. The main nominal parameters of the structures considered for both measurements and calculation are summarized in Table I. All samples are grown on top of a high-resistivity p-type Si substrate by means of molecular beam epitaxy. The virtual substrate consists of a graded buffer with a slope of 20% Ge/µm. The Hall mobility measurements have been performed using either standard Van der Pauw or differential Hall techniques [8].

The gated structure simulated to study in sub-section III.5 the gate effect on the electron transport is not the subject of comparison with experiments.

### B. NOMINAL AND EFFECTIVE DOPING PROFILES

One of the aims of the study is to validate our modeling approach by comparing the calculations with experimental data. To make relevant the comparison we must ensure that the simulated doping profiles are close to the actual ones. Indeed, each sample includes pulse-doped layers characterized by nominal width and doping level (see Table I). In practice, a quite important difference may occur between actual activated dopant concentrations and nominal values which, in addition, do not consider any effect of dopant diffusion. In other words, a structure may have a doping profile very different from expected. It has been estimated that in some cases the uncertainty on the doping levels can reach a factor of 2. Such uncertainty is not acceptable if we consider that the impurity scattering rate is proportional to doping density, as stated in Eq. (8). A correction of the effective profile is then required in most cases for a useful comparison with measurements.

In the ungated structures considered in this section, the carrier density is directly correlated to the impurity concentration. As initial guess, the nominal density can be roughly estimated from the width and nominal doping levels of the supply layers. The first step of the correction procedure consists in comparing nominal and measured densities. A strong discrepancy is a clear sign that a correction is required. The measured density is then to be compared with the density calculated by solving self-consistently the Poisson and Schrödinger equations for a nominal doping profile.

The procedure starts at low temperature (T = 77 K), i.e. when most free carriers reside in the main channel, as shown in Fig. 2. The width, the doping level and profile of supply layers are then empirically adjusted to get a simulated electron density equal to the measured one. We proceed in the same way at 300 K to refine the doping profile. As shown in Fig. 2, for a given doping profile the carrier density in the strained channel is almost the same at 300 K and 77 K. The temperature only influences the carrier density in the supply layers and their surroundings. Thus the adjustment of the doping profile at room temperature must be made to fit the measured total density without changing the density in the strained-Si channel.

C. MOBILITY RESULTS: COMPARISON WITH MEASUREMENTS

The drift mobility is computed under uniform in-plane electric field of 0.5 kV/cm at room temperature. Measured and calculated mobilities are displayed in Fig. 3 as a function of the measured (open symbols) and calculated (full symbols) electron density, respectively. The mobility depends obviously on several technological parameters and cannot be a unique function of the electron density. It is however a common representation showing the main trend: the mobility roughly decreases as the density increases, which is related to higher doping of supply layers and thus higher impurity scattering. But it is shown in next sub-section that the spacer thickness may have a stronger influence on the mobility than the doping density itself.

The Hall scattering factor for electrons in a 2DEG was shown to be very close to unity in a large range of temperatures [27]. Assuming this result is unchanged for strained material makes possible the direct comparison of our calculated density and drift mobility with Hall measurements. This assumption, however, may fail for electrons in SiGe, which can be a possible source of error in the case of structure having a strong parasitic path.

As shown in Fig. 3, for each simulated structure a fairly good agreement is found with measurements on a wide range of electron density. At very low density, the maximum calculated mobility reaches 2800 cm$^2$/Vs. In this structure the supply doping is small and remote enough to make the impurity scattering negligible. The electrons are confined in the strained-Si channel and the predominant scattering mechanisms are the electron-phonon interactions. This calculated mobility matches very well the value measured by differential Hall technique (2700 cm$^2$/Vs, open square in Fig. 3) and the best other values reported at low electron density [7]. This agreement validates our approach of phonon scattering modeling. The increase of quantum well thickness reduces the level spacing and affects the phonon scattering rates but it results in a limited effect on the peak mobility. For QW larger than 9 nm the phonon-limited mobility does not exceed 3000 cm$^2$/Vs. It should be noted that using the same phonon coupling parameters, the mobility calculated in a 2DEG is slightly lower than that calculated by neglecting quantization effects (3250 cm$^2$/Vs) [11]. It is consistent with the calculated phonon scattering rates: in the energy range 0-50 meV, for each electron-phonon process the scattering rate is smaller in a 3DEG than in a 2DEG [13,14]. With a different set of phonon parameters Yamada et al. calculated a phonon-limited mobility of 4000 cm$^2$/Vs for a 3DEG formed in strained Si [28]. Bufler et al. obtained a mobility of 2250 cm$^2$/Vs using full-band Monte Carlo simulation [29]. A better agreement with experimental results is found using our model.

For moderate densities, few carriers are in the parasitic paths, and the decreasing mobility is essentially due to the higher remote impurity scattering. When increasing the doping levels, the parasitic channels form deeper quantum wells and more carriers contribute to the parasitic conduction (up to 40% in some cases). At high electron density, the mobility is then dominated by remote impurities for the electrons in the strained channel and by background impurities for the electrons bound to the parasitic paths. It is remarkable in this case that the mobility never falls off under 1100 cm$^2$/Vs. The good agreement between measurement and simulation on the full range of carrier density may be considered as a validation of our treatment of impurity scattering including a simple approach of the description of screening effects.

D. INFLUENCE OF THE SPACER THICKNESS

The influence of remote impurities is partially controlled by the width of the spacer layers that strongly influences the exponential term in Eq. (8). At given supply doping, increasing the spacer thickness reduces the impurity scattering rate for electrons in the main channel but also reduces the electron transfer from the supply layers to the channel. The current in the structure results from the balance between these two effects.

This is illustrated in Fig. 4 that shows the electrons density and mobility as a function of the top spacer thickness in a typical DSD structure, i.e. the C2766 sample (see Table I). In this case, increasing the spacer thickness from 1 nm to 5 nm causes a small reduction in the total electron density (from $5.6 \times 10^{12}$ cm$^{-2}$ to $5.3 \times 10^{12}$ cm$^{-2}$) but yields also a significant increase in mobility (from 1180 cm$^2$/Vs to 1540 cm$^2$/Vs). This clearly confirms that the mobility in this type of structure is not a unique function of the carrier density.

It should be noted that this increase in mobility is obtained in spite of a detrimental enhancement of the fraction of electron density in the parasitic paths in which the transport properties are poor. As shown in Fig. 5, the fraction of electrons in the main strained-Si channel decreases from 93% to 63% as $w_{sp_1}$ increases up to 5 nm.

The strong effect of impurity scattering on the electron transport in this structure is clearly shown in Fig. 6 where we plot the mobility calculated as a function of $w_{sp_1}$ depending on whether the impurity scattering mechanism is included (solid line) or not (dashed line). If impurity scattering is not taken into account the mobility tends to decrease as $w_{sp_1}$ increases. This is only resulting from an increase of the overlap integral in the phonon scattering rates given in Eqs (1), (2) and (3). The introduction of remote impurity scattering into the algorithm reduces significantly the mobility but as $w_{sp_1}$ increases the reduction in impurity scattering rate dominates the increase in phonon scattering rate, which results in a higher overall mobility.

E. INFLUENCE OF THE BIAS IN A MODFET STRUCTURE

We now consider a gated device whose vertical architecture is of the same type as the DSD structure schematized in Fig. 1, with a Schottky-gate on the top. The technological parameters are very similar to that of sample C2309 but optimized for MODFET operation: $W_{ch} = 9$ nm, $W_{SiCap} = 4$ nm, $W_1 = 3.6$ nm $W_{sp_1} = 3.6$ nm, $W_{sp_2} = 3$ nm, $W_{\delta_1} = 3.4$ nm, $W_{\delta_2} = 4$ nm, $N_{\delta_1} = 1.5 \times 10^{19}$cm$^{-3}$, $N_{\delta_2} = 4 \times 10^{18}$cm$^{-3}$. In such a gated structure the conducting electron density is no more controlled by the doping density in the supply layers but by the gate field effect, which modifies the influence of remote impurity scattering.

With a Schottky-gate in normal operation the top δ-doped layer is fully depleted and most carriers stand in the strained channel. Hence, in the frame of our impurity screening model, the first subband of the normal valleys is obviously the subband that should be taken into account in the evaluation of the screening function from Eq. (9). The influence of gate voltage $V_{GS}$ on the calculated electron mobility is shown in Fig. 7. As the gate voltage increases, the mobility tends to increase and then to saturate near 2140 cm$^2$/Vs for positive gate bias. The rising part of this curve can be easily explained: it is due to the gate-induced enhancement of density $n_1$ on the lowest subband, which increases the screening function $Q_{scr}$ (Eq. (9)) and reduces the impurity scattering rate (Eq. (8)). The mobility saturation at high gate bias, however, is somewhat unexpected. Consistently, the intra subband impurity scattering rate for the first level of the normal valleys exhibits a similar plateau in Fig. 8. To understand this behavior, we also plot in Fig. 8 the density on this level as a function of $V_{GS}$. A significant increase in the slope of $n_1$ is observed for voltages larger than 0 V, which is a sign of a significant change in the charge-control operation resulting in a higher gate capacitance. This observation is confirmed in Fig. 9 where the electron concentration profile in the channel is shown for different gate voltages. While the maximum electron concentration stand on the right side (bottom) of the channel at low gate bias, for $V_{GS}$ greater than 0.1 V this maximum concentration shifts to the left side of the channel. This reduces the effective distance between the electron charge and the gate, hence increasing the charge control capacitance. This space change of the maximum electron concentration with the bias increases the numerator of the integrand in the impurity scattering rate given by Eq. (8), which tends to compensate the increase of the screening function $Q_{scr}$ in the denominator. It is the origin of the plateau in the scattering rate observed in Fig. 8, which is directly reflected on the mobility in Fig. 7.

It should be noted that whatever the bias voltage, the fraction of carriers contributing to a parasitic conduction remains low and reaches only 4% of the total electron density at $V_{GS}$ = 0.4 V. It explains that the mobility is always greater than in the ungated modulation doped structure having a similar layer stack (sample C2309, see Fig. 3).

## IV. CONCLUSION

We have computed the electron mobility in 2DEG formed in tensile-strained Si/SiGe modulation doped structures by means of Monte Carlo simulation. To make relevant the direct comparison with experimental data, we have empirically modified the nominal doping profile in simulated structures. In ungated structures, a good agreement is found with Hall measurements on a wide range of doping conditions and carrier densities, which demonstrates the correctness of the scattering models used. The mobility typically varies between 1100 cm$^2$/Vs and 2800 cm$^2$/Vs. It is shown to be significantly influenced by the thickness of the spacer layer separating the strained-Si channel from the delta-doped supply layers.

In a gated structure designed for MODFET operation, the contribution of conducting parasitic paths becomes negligible and the mobility is higher than in comparable ungated samples. Additionally, the variation of gate bias $V_{GS}$ modulates the screening of remote impurities by free carriers. The most important consequence is in an increase in mobility as $V_{GS}$ increases.

## ACKNOWLEDGEMENTS

A part of this work was supported by the European Community under the IST Program (SIGMUND Project).

## REFERENCES

[1] C.G. Van de Walle, in *Properties of strained and relaxed silicon germanium*, EMIS Datareviews Series 12, edited by E. Kasper (INSPEC, London, 1995), pp. 94-102.

[2] S. Galdin, P. Dollfus, V. Aubry-Fortuna, P. Hesto, H.J. Osten, Semicond. Sci. Technol. 15, 565-572 (2000).

[3] M. Zeuner, T. Hackbarth, M. Enciso-Aguilar, F. Aniel, and H. von Känel, Jpn. J. Appl. Phys. 42, 2363-2365 (2003).

[4] F. Aniel, M. Enciso-Aguilar, L. Giguerre, P. Crozat, R. Adde, T. Mack, U. Seiler, T. Hackbarth, H. J. Herzog, U. König and B. Raynor, Solid-State Electron. 47, 283-289 (2003).

[5] N. Sugii, D. Hisamoto, K. Washio, N. Yokoyama and S. Kimura, IEDM 2001 Tech. Dig. 737-740 (2001).

[6] F.M. Bufler and W. Fichtner, IEEE Trans. Electron Dev. 50, 278-284 (2003).

[7] K. Ismail, S.F. Nelson, J.O. Chu, and B.S. Meyerson, Appl. Phys. Lett. 63, 660-662 (1993).

[8] T. Hackbarth, G. Hoeck, H. J. Herzog, and M. Zeuner, J. Cryst. Growth 201/202, 734-738 (1999).

[9] P. Dollfus, S. Galdin, H. J. Osten, and P. Hesto, J. Mat. Sci : Mat. in Elec. 12, 245-248 (2001).

[10] S. Richard, .N Cavassilas, F. Aniel, G. Fishman, J. Appl. Phys. 94, 5088-5094 (2003).

[11] P. Dollfus, J. Appl. Phys. 82, 3911-3916 (1997).

[12] D. K. Ferry, Phys. Rev. B 14, 1605 (1976).

[13] M. V. Fischetti and S. E. Laux, Phys. Rev. B 48, 2244 (1993).

[14] F. Monsef, P. Dollfus, S. Galdin, and A. Bournel, Phys. Rev. B 65, 212304 (2002) – Phys. Rev. B 67, 059903(E) (2003).

[15] J. W. Harrison, Phys. Rev. B 13, 5347 (1976).

[16] V. Venkataraman, C. W. Lin, and J. C. Sturm, J. Appl. Phys. 73, 7427 (1993).

[17] T. Ando, A. B. Fowler, and F. Stern, Rev. Mod. Phys. 54, 437 (1982).

[18] F. Stern, Phys. Rev. Lett. 44, 1469 (1980).

[19] K. Yokoyama and K. Hess, Phys. Rev. B 33, 5595 (1986).

[20] J.M. Ziman, *Principles of the theory of solids*, 2nd Edition (Cambridge University Press, Cambridge, 1972).

[21] E. D. Siggia and P. C. Kwok, Phys. Rev. B 2, 1024 (1970).

[22] F. Gamiz, J. A. Lopez-Villanueva, J. A. Jimenez-Tejada, I. Melchor, and A. Palma, J. Appl. Phys. 75, 924 (1994).

[23] J. L. Farvacque, Phys. Rev. B 67, 195324 (2003).

[24] D.K. Ferry and S.M. Goodnick, *Transport in nanostructures* (Cambridge University Press, Cambridge, 1997).

[25] H. Kosina, Phys. Stat. Sol. (a) 163, 475 (1997).

[26] A. Fetter, Phys. Rev. B 10, 3739 (1974).

[27] J. Jungemann, D. Dundenbostel, and B. Meinerzhagen, IEEE Trans. Elec. Dev. ED-46, 1803 (1999).

[28] T. Yamada, J.R. Zhou, H. Miyata, and D.K. Ferry, IEEE Trans. Elec. Dev. ED-41, 1513 (1994).

[29] F.M. Bufler and W. Fichtner, Appl. Phys. Lett. 81, 82 (2002).

Monsef et al. **Table I**

Table I. Nominal technological parameters of $Si_{1-x}Ge_x/Si/Si_{1-x}Ge_x$ modulation-doped structures investigated both experimentally and theoretically. The thickness symbols are defined in Fig. 1. $N_{\delta_1}$ and $N_{\delta_2}$ are the doping levels in the top and bottom supply layers, respectively.

| Sample N° | $x$ (%) | $W_{cap}$ (nm) | $W_1$ (nm) | $W_{ch}$ (nm) | $W_{sp_1}$ (nm) | $W_{\delta_1}$ (nm) | $N_{\delta_1}$ (cm$^{-3}$) | $W_{sp_2}$ (nm) | $W_{\delta_2}$ (nm) | $N_{\delta_2}$ (cm$^{-3}$) |
|---|---|---|---|---|---|---|---|---|---|---|
| C2311 | 35 | 4 | 20 | 9 | 15 | 10 | $5\times10^{18}$ | | | |
| C2014 | 45 | 4 | 7 | 9 | 6 | 4 | $1.5\times10^{19}$ | 6 | 4 | $4\times10^{18}$ |
| C2309 | 45 | 3 | 7 | 9 | 3.1 | 3.6 | $1.5\times10^{19}$ | 3.1 | 4 | $4\times10^{18}$ |
| C2589 | 40 | 4 | 8 | 9 | 3 | 4 | $1.5\times10^{19}$ | 4 | 5 | $2.4\times10^{18}$ |
| C1898 | 45 | 4 | 8 | 9 | 3 | 4 | $1.5\times10^{19}$ | 3 | 4 | $4\times10^{18}$ |
| C2766 | 45 | 3 | 8 | 9 | 4.5 | 5 | $1.5\times10^{19}$ | 3 | 5 | $4\times10^{18}$ |
| C2720 | 40 | 3 | 8 | 9 | 3.5 | 5 | $1.5\times10^{19}$ | 4 | 5 | $4\times10^{18}$ |

FIGURE CAPTIONS

Figure 1. Schematic cross-section of n-type modulation-doped structures with either a single-side pulse-doped layer (SSD) or a double-side pulse-doped layer (DSD).

Figure 2. Electron concentration in the sample C2766 calculated with nominal doping profile at 77 K and 300 K.

Figure 3. Experimental (open symbols) and calculated (full symbols) mobility for various samples as a function of the total electron density. Three values of germanium content $x$ are considered: $x = 45\%$ (circles), $x = 40\%$ (triangles), $x = 35\%$ (diamonds).

Figure 4. Total electron density and mobility as a function of the top spacer width $W_{sp_1}$, all other things being equal. $W_{sp_1} = 4.5$ nm corresponds to sample C2766.

Figure 5. Distribution of electron density in the different paths as a function of the top spacer width $W_{sp_1}$.

Figure 6. Influence of the top spacer width $W_{sp_1}$ on the mobility depending on whether the impurity scattering is included or not.

Figure 7. Electron mobility in the gated modulation-doped structure as a function of gate voltage.

Figure 8. Intra-subband impurity scattering rate and density on the first level of the normal valleys as a function of gate voltage.

Figure 9. Profile of electron concentration in the strained-Si channel for various gate voltages. The vertical dashed lines indicate the SiGe/Si interfaces. The maximum concentration shifts from the bottom interface at low gate bias to the top interface at high gate bias.

| Si cap (n.i.d.) | $W_{SiCap}$ | Si cap (n.i.d.) |
| --- | --- | --- |
| SiGe (n.i.d.) | $W_1$ | SiGe (n.i.d.) |
| n-doped SiGe | $W_{\delta_1}$ | n-doped SiGe |
| SiGe spacer (n.i.d.) | $W_{sp_1}$ | SiGe spacer (n.i.d.) |
| Si channel (n.i.d.) | $W_{ch}$ | Si channel (n.i.d.) |
| SiGe spacer (n.i.d.) | $W_{sp_2}$ | |
| n-doped SiGe | $W_{\delta_2}$ | |
| SiGe buffer (n.i.d.) | | SiGe buffer (n.i.d.) |
| DSD structure | | SSD structure |

Monsef et al.  **Figure 2**

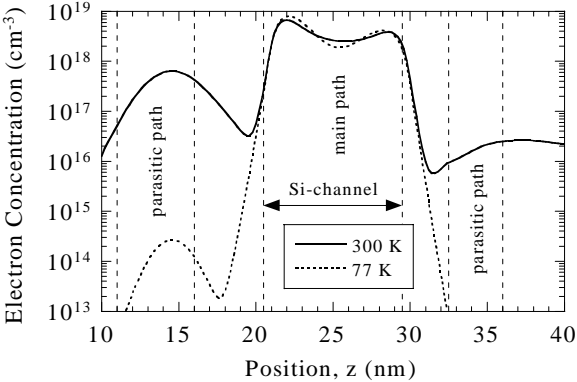

Monsef et al.        **Figure 3**

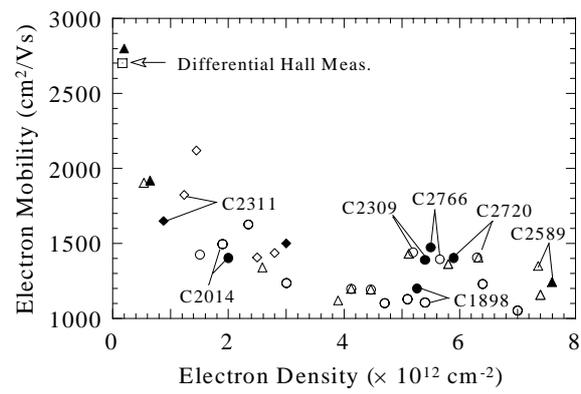

Monsef et al. **Figure 4**

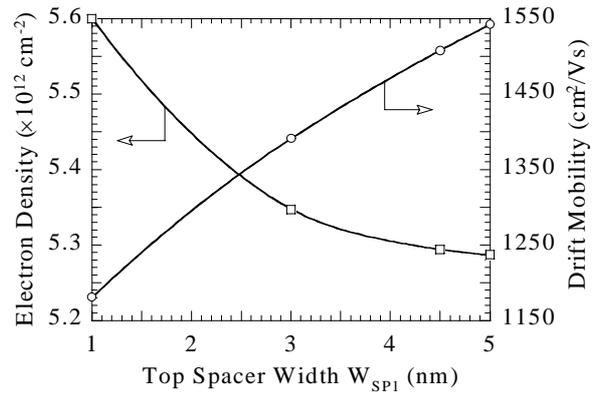

Monsef et al.     **Figure 5**

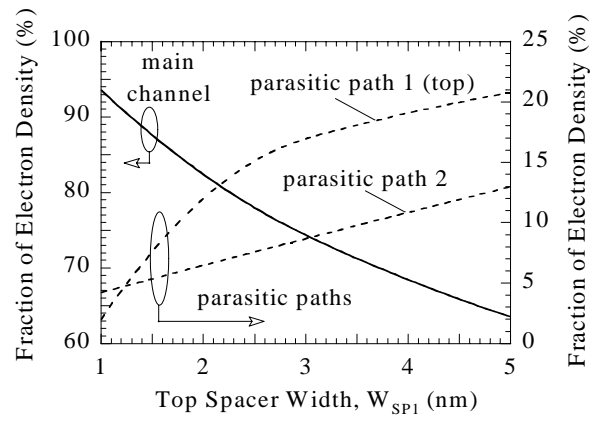

Monsef et al.     **Figure 6**
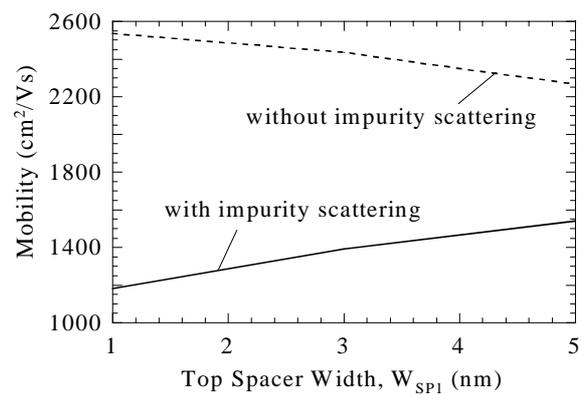

Monsef et al.     **Figure 7**

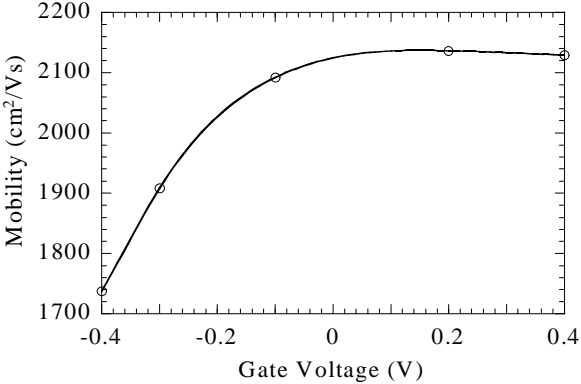

Monsef et al.     **Figure 8**

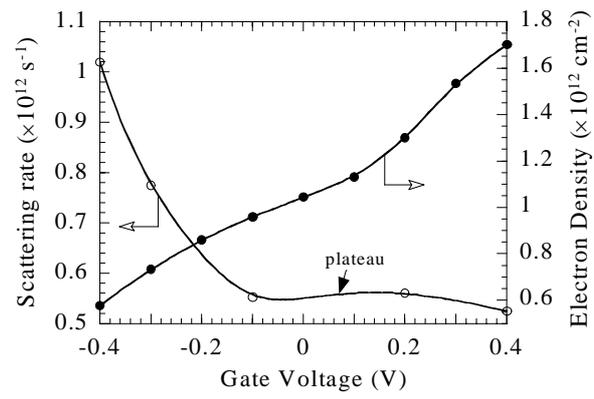

Monsef et al.     **Figure 9**

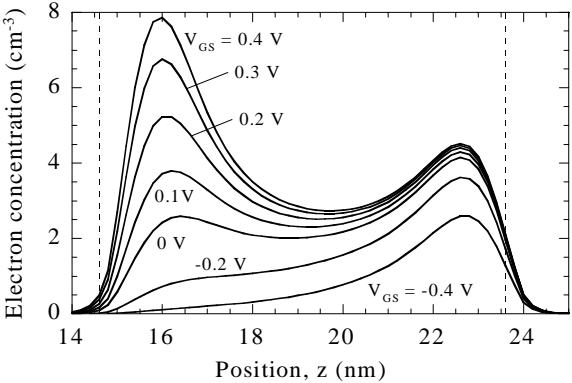